# Dual Metal-Gate Planar Field-Effect Transistor for Electrostatically Doped CMOS Design

Tillmann Krauss, Frank Wessely and Udo Schwalke
Institute for Semiconductor Technology and Nanoelectronics
Technische Universität Darmstadt
Darmstadt, Germany
krauss@iht.tu-darmstadt.de

*Abstract*— In this paper, we demonstrate by simulation the general usability of an electrostatically doped and electrically reconfigurable planar field-effect transistor (FET) structure. The device concept is partly based on our already published and fabricated silicon nanowire 3D-FET technology. The technological key features of this general purpose FET contain Schottky S/D junctions on a silicon-on-insulator (SOI) substrate additionally enabling high-temperature operation of the proposed device. The transistors charge carrier type, i.e. n- or p-type FET, is electrically switchable on the fly by applying a control-gate voltage, which significantly increases the flexibility and versatility in digital integrated circuits.

*Keywords—voltage-programmable; ambipolar; reconfigurable; electrostatic doping; FET; SOI*

## I. Introduction

In today's aggressively scaled MOSFET devices, the dominating leakage path for OFF-state currents originate from PN-junction and bulk-leakage [1], [2]. Unfortunately, these leakage currents increase with temperature limiting the maximum operation temperature of modern silicon MOSFETs. SOI based FET technologies can reduce bulk-leakage currents significantly and increase operation temperatures but PN-junction leakage remains present even in SOI FETs. Therefore, our device structure features Schottky-barrier source/drain contacts instead of PN-junctions adding to the high temperature robustness as shown previously in refs. [3], [4] and [5].

The intrinsic fusion of an ambipolar device characteristic with a separated current flow control-gate leads to a extraordinary leakage current reduction and high on-to-off current ratio. Furthermore, the flexibility to select n-channel and p-channel operation on the fly paves the way for reconfigurable integrated circuits designs [10]. Finally yet importantly, as no standard CMOS doping process steps are required during manufacturing, the device performance does not deteriorate due to statistic dopant fluctuation in ultra-scaled devices and dopant dependent reduction of carrier mobility that typically arise in modern MOSFET devices.

To the best of our knowledge, all similar devices with an ambipolar behavior and a separated electrode for current flow control are based on nanowire structures as, for example, carbon nanotubes or silicon nanowires. A summary of these devices is given in ref. [6] but we want to point out that the here discussed planar device structure is quite different.

## II. Device Structure

A cross-section of the planar device structure is depicted in Fig. 1. It is partly based on our published SiNW FET technology as reported in refs. [3], [4], [5]. Our new concept, as well as the previous SiNW technology, is experimentally based on a virtually dopant-free SOI substrate, i.e. with the lowest commercially available boron background doping of $2\times10^{15}$ cm$^{-3}$ [7].

The simulated structure is based on a SOI substrate with a top silicon layer of 40 nm and a buried silicon oxide layer (BOX) of 50 nm. It features three gate electrodes, namely front-gate (FG) M1, FG M2 and back-gate (BG). It is important that FG M1 and FG M2 are electrically shorted but nevertheless can differ in metal work function. The support wafer or handle silicon layer serves as the BG contact and is insulated by the BOX from the top layer. In the center of the top layer, the electrodes FG M1 and M2 are placed into a recess with 100 nm gate length each or 200 nm total ($L_{FG}$ in Fig. 1). A hafnium oxide layer of 10 nm thickness realizes the insulation of the FG electrodes and the channel height beneath the FG region is 10 nm if not mentioned otherwise. Finally, the source and drain contacts are connected to the sides of the top silicon layer with a lateral distance of 1 µm ($L_{BG}$ in Fig. 1).

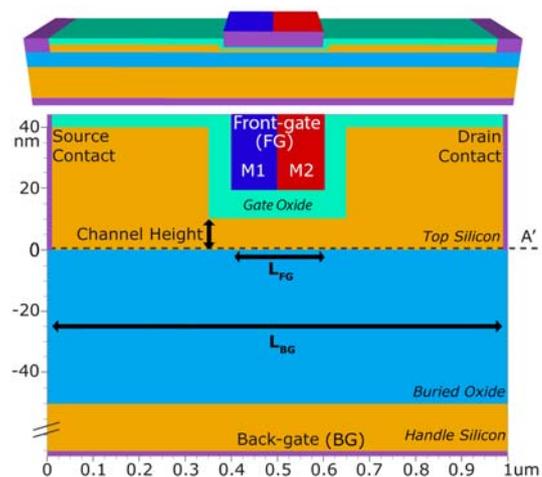

Fig. 1. Cross-section of the simulated planar dual metal front-gate FET.

## III. RESULTS AND DISCUSSION

In this chapter, we will simulate the effects of biasing conditions, physical geometry modifications as well as FG metal work function on the electrical performance of the proposed device. The device is simulated with Silvaco Atlas and its drift-diffusion and universal Schottky transport models.

In order to ease the understating of the device operation modes, the FET can be regarded as an intrinsic combination of two interacting transistors. The two gate electrodes, FG and BG respectively, represent these transistors.

The BG electrode forms the first transistor and influences the whole channel region from source to drain as an ambipolar enhancement mode or normally-off FET ($L_{BG}$ in Fig. 1). This transistor defines the type of FET, i.e. n- or p-type. On the other hand, the second transistor formed by the FG electrodes influences only the center of the channel ($L_{FG}$ in Fig. 1). The FG transistor resembles a depletion mode or normally-on FET and affects the channel built by the BG transistor between source and drain. The combination of these two operation modes can be termed as dehancement mode operation.

### A. Back-Gate Enhancement Mode Operation

Fig. 2 depicts a BG voltage ($V_{BG}$) sweep with an electrically floating FG and a source bias ($V_S$) 100 mV. The sweep clearly shows the ambipolar nature of the BG transistor. A sufficiently high BG bias voltage results in an increasing of source current as either holes ($V_{BG} < 0.5$ V) or electrons ($V_{BG} > 1$ V) are forming a channel at the top-silicon-to-BOX interface.

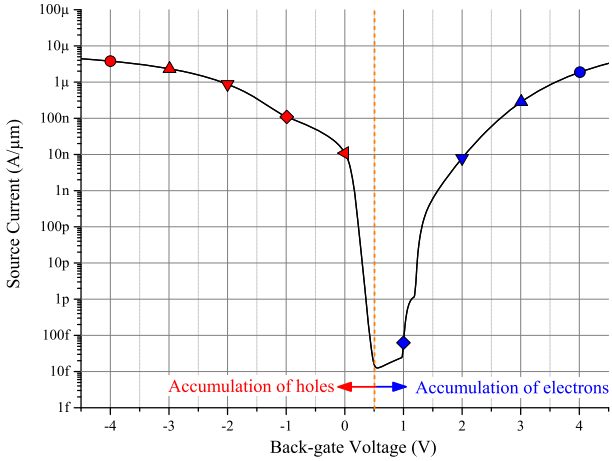

Fig. 2. Voltage sweep of the back-gate contact with floating front-gate condition at a source voltage of $V_S = 0.1$ V. Symbol correspond to operating points in Fig. 6.

In Fig. 3 and 4 two BG biasing conditions for a single metal front-gate with a work function of 4.7 eV are depicted as band diagrams. For a positive BG voltage, electrons get attracted and accumulate at the BOX interface (Fig. 3). In contrast, an attraction of holes is realized with a sufficiently negative BG voltage (Fig. 4).

The attracted charge carriers originate from the mid-gap Schottky barrier source/drain contacts with a work function of 4.6 eV. The electrical field of the BG alters the shape especially width of the Schottky barriers and therefore influences the probability for tunneling and thermionic field emission of charge carriers through the barriers.

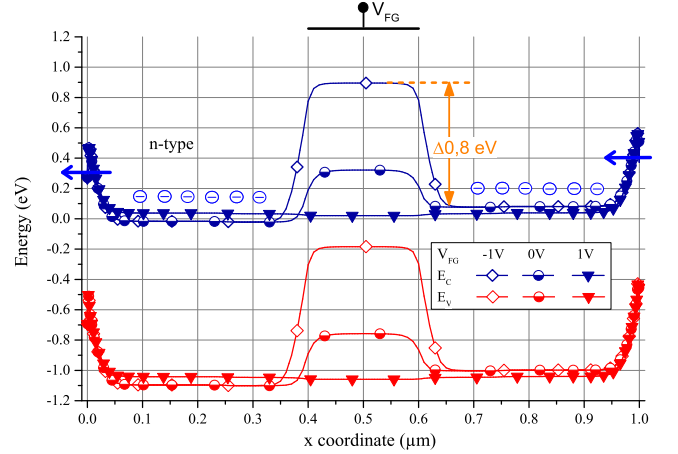

Fig. 3. Band diagram for $V_{BG} = +3$ V, $V_S = 0.1$ V for a single metal front-gate (M1 = M2) with a work function of 4.7 eV. Filled symbols represent on-state transistor, partly filled symbols represent transition state. Open symbols visualise off-state.

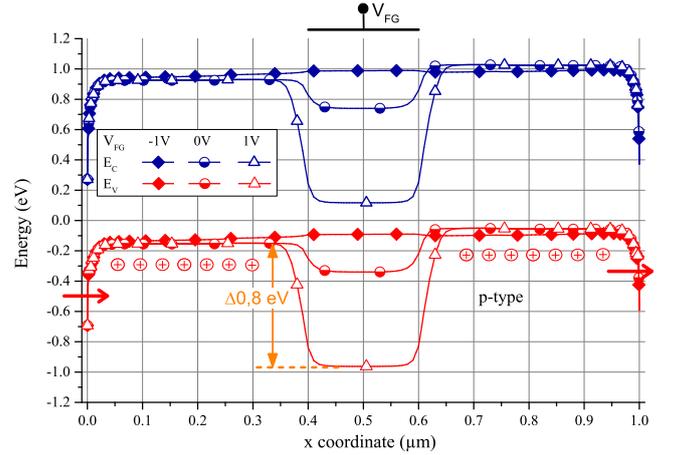

Fig. 4. Band diagram for $V_{BG} = -3$ V, $V_S = 0.1$ V for a single metal front-gate (M1 = M2) with a work function of 4.7 eV. Filled symbols represent on-state transistor, partly filled symbols represent transition state. Open symbols visualise off-state.

The physical parameters for the Schottky barrier simulation are estimated by fitting simulated and measured back-gate voltage sweeps of fabricated SiNW FET devices with 70 nm x 70 nm rectangular diameter and 50um length (Fig. 5) [3], [4], [5]. Deviations between measurement and simulation are mainly caused by not modeled complex Schottky barrier effects (not shown here) as, for example, barrier lowering due to image charges.

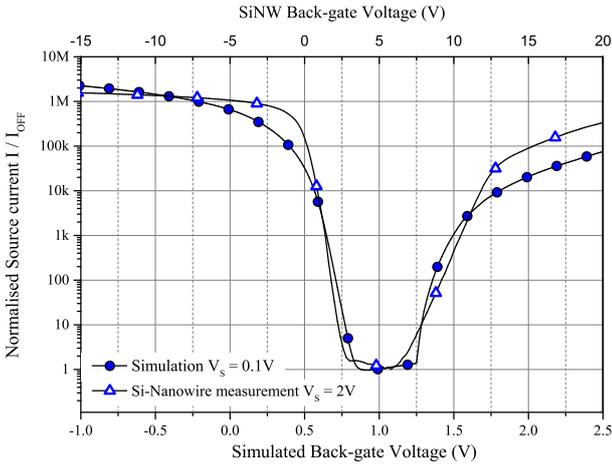

Fig. 5. Comparison of simulated and measured back-gate voltage sweeps of silicon nanowire FET with electrically floating front-gate condition.

### B. Front-Gate Depletion Mode Operation

The FG electrode influences the accumulated (i.e. electrostatically doped) channel of the BG. Therefore, the FG transistor resembles a depletion mode FET.

The input characteristics for a single metal FG FET for various BG biasing conditions are shown in Fig. 6. The simulation shows a remarkable source/drain leakage current of approximately 1 aA/µm and an extraordinary on-to-off current ratio of up to ~12 decades. The low leakage currents are the result of the high potential barriers for electrons and holes of up to 0.8 eV formed by the FG as can be seen in the band diagrams in Fig. 3 and Fig. 4. As expected, the on-current on the other hand is directly linked to the magnitude of back-gate voltage.

Furthermore, the threshold voltage can be shifted by approximately 100 mV/$V_{BG}$ while only slightly degrading the subthreshold slope by 0.5 mV/dec/$V_{BG}$ granting additional flexibility to circuit designers (Fig. 6).

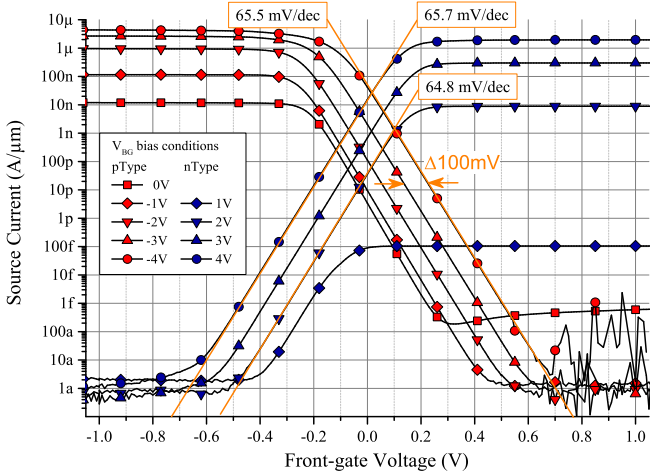

Fig. 6. Voltage sweeps for single metal front-gate with a work function of 4.7 eV at $V_S = 0.1$ V for various back-gate voltages ($V_{BG} = -4…+4$ V). Symbol shapes correspond to operating points in Fig. 2.

### C. Effect of Front-Gate Length and Gate Oxide Permittivity

The electrical effect of the FG is mainly determined by the FG length and effective electric permittivity of the material stack between the accumulated BG channel and the FG metal electrode.

As already known from other FET structures, a reduction of the gate length ($L_{FG}$) considerably degrades the subthreshold slope and increases the threshold voltage. As expected, using a high-k oxide like hafnium oxide can mitigate this degradation as can be seen in Fig. 8 for $L_{FG}$ 10 and 20 nm and hafnium oxide (Fig. 7, open symbols).

Special to our device is the missing direct interface between the FG oxide and the channel. Additionally, no high temperature processing steps are required after FG oxide formation. Therefore, there are no technological handicaps like interface deterioration and temperature instabilities of high-k insulators which are known to degrade the device performance when high-k gate oxides like gadolinium oxide or other lanthanide oxides are introduced, as reported in references [8] and [9] for example.

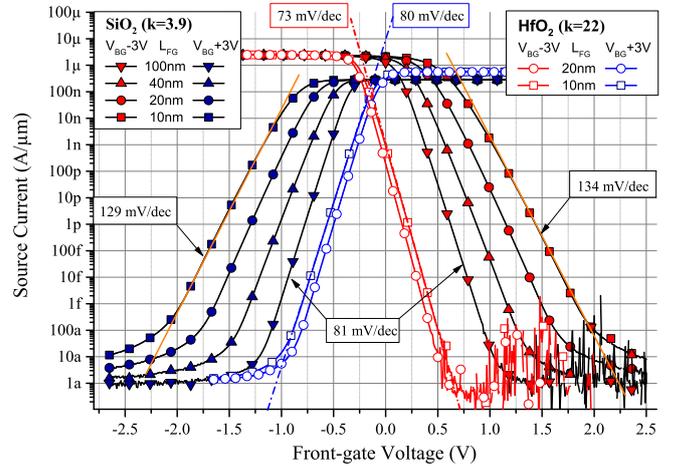

Fig. 7. Effect of front-gate length and oxide permittivity on subthreshold slope and theshold voltage for a single metal front-gate with a work function of 4.6 eV (M1 = M2) at $V_S = 0.1$ V.

### D. Effect of Dual Metal Front-Gate

Besides the already discussed parameters, the metal work function of the FG electrode is a technological key feature as it is directly linked to the threshold voltage characteristic of field-effect transistors.

The impact of different work functions on the threshold voltage for a single metal FG with $L_{FG}$ 200 nm is depicted in Fig. 8. A work function difference of 0.6 eV results in one-to-one shifts of threshold voltages of 0.6 V for p- (red symbols) and n-type (blue symbols) operation mode (Fig. 8). With the aim to reduce switching currents in CMOS designs it is obligatory to technologically shift the threshold voltages discretely for n- and p-type transistors. An arrangement of two metals with different work functions in a dual metal FG can fulfill this requirement as shown in Fig. 9.

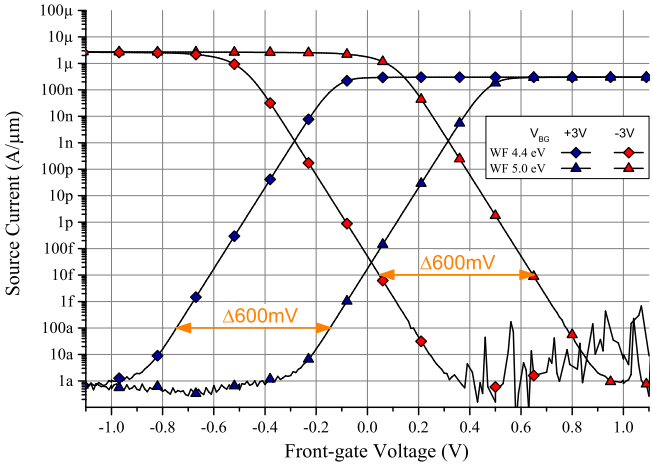

Fig. 8. Effect of work function on threshold voltage for single metal front-gate transistor ($L_{FG}$ 200 nm) with metal work functions of 4.4 eV and 5.0 eV at $V_S = 0.1$ V.

The combination of two metal work functions in a dual metal FG decreases the crossing current for a FG voltage of 20 mV from 1 nA/µm for a single metal FG with 4.7 eV down to 100 fA/µm for a dual metal FG with work functions of 4.4 eV for M1 and 5 eV for M2 (Fig. 9).

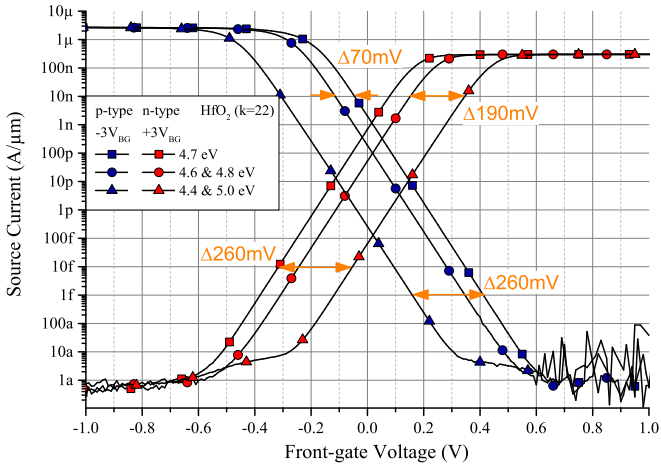

Fig. 9. Effect of dual metal front-gate ($L_{FG}$ 2x100 nm) metal work function difference on threshold voltage.

As the effective FG length is reduced from 1x 200 nm to 2x 100 nm ($L_{FG}$ M1 = $L_{FG}$ M2 = 100 nm) the threshold voltage shift is not an one-to-one relation to metal work function differences. In order to visualize the origin of the threshold voltage shift, band diagrams are depicted in Fig. 10 and Fig. 11. For a FG voltage of 0 V (half-filled symbols) only the potential barrier of either M1 or M2 is interacting with the carrier flow through the channel.

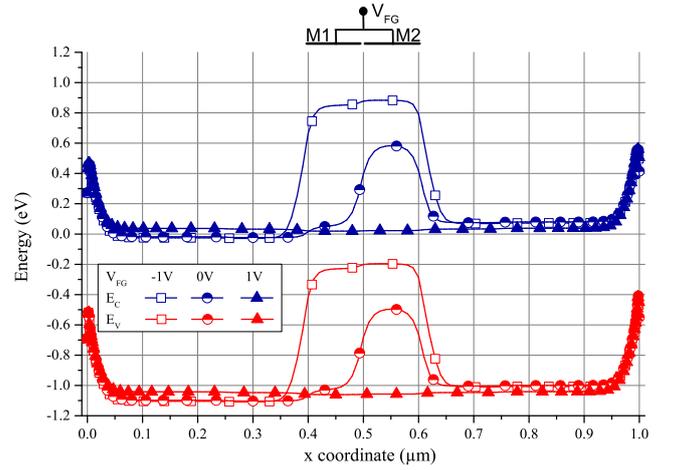

Fig. 10. Band diagram for positive back-gate biasing $V_{BG} = +3$ V at a source voltage of $V_S = 0.1$ V for a dual metal front-gate with work functions M1 = 4.4 eV and M2 = 5.0 eV. Filled symbols show on-state transistor, half-filled symbols represent transition state. Open symbols visualise off-state.

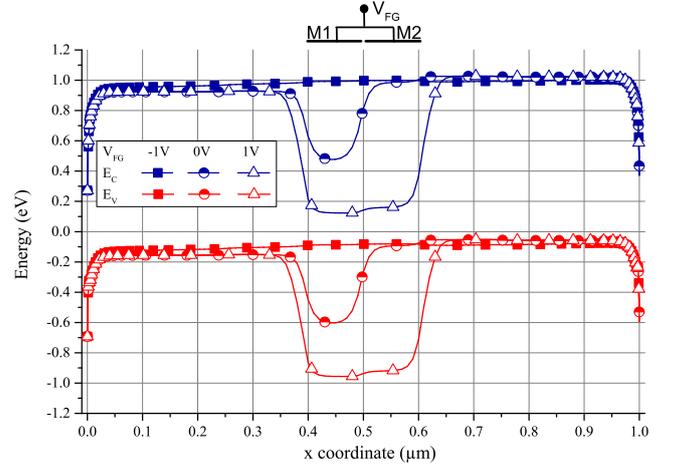

Fig. 11. Band diagram for negative back-gate biasing $V_{BG} = -3$ V at a source voltage of $V_S = 0.1$ V for dual metal front-gate with metal work functions M1 = 4.4 eV and M2 = 5.0 eV. Filled symbols show on-state transistor, half-filled symbols represent transition state. Open symbols visualise off-state.

*E. CMOS Inverter*

In this chapter, we will demonstrate the general usability of the dual metal FG device in reconfigurable digital circuits. For this purpose, we selected the classic CMOS inverter, as it is one of the most basic building blocks for digital designs (Fig. 13).

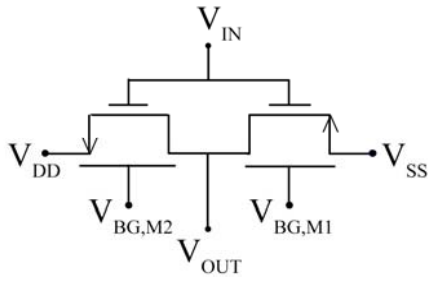

Fig. 12. Basic CMOS Inverter circuit based on dehancement mode FET.

The output characteristics of a CMOS inverter with a dual metal FG for an asymmetric supply voltage ($V_{SS}$ = 0.5 V, $V_{DD}$ = 0 V, M1 = 4.2 eV, M2 = 5.2 eV) and symmetric supply voltage ($V_{SS}$ = 0.2 V, $V_{DD}$ = -0.2 V, M1 = 4.4 eV, M2 = 5.0 eV) are depicted in Fig. 14 and Fig. 15. As can be seen, higher BG voltages tend to degrade the inverter performance as the threshold voltage difference between n- and p-type FET increases resulting in a surge of cross-current (CC) and rounding as well as slope reduction of the transfer characteristic (TC) (see also Fig. 2). Vice versa, a larger work function difference enhances the inverter performance as the threshold voltage difference is reduced.

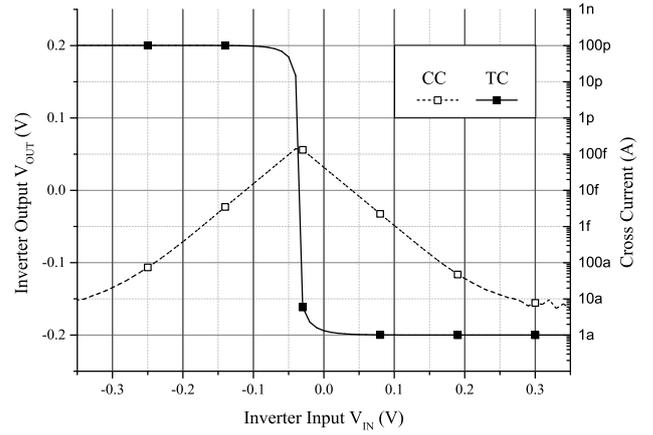

Fig. 14. Inverter transfer characteristic (TC) and cross current (CC) for dual metal front-gate device. Work function of FG metal M1 = 4.4 and M2 = 5 eV. Symmetric supply $V_{SS}$ = 0.2 V, $V_{DD}$ = -0.2 V.

## IV. CONCLUSION

In conclusion, the simulated dual metal front-gate device with midgap Schottky-barrier source and drain contacts can serve as a versatile, reconfigurable building block for logic circuit designs. The dual metal front-gate device demonstrates a noteworthy low off-state leakage current and adjustable threshold voltage, which is a key aspect of low power silicon CMOS designs. The ability to electrically select the transistor type (i.e. P- or NFET) on the fly, permits digital circuit designers to build switchable NAND/NOR cells with the very same physical transistors in place saving precious silicon area [10]. Using unconventional, innovative logic synthesis methodologies like MIXSyn the total transistor count for logic applications can be further reduced [11], [12].

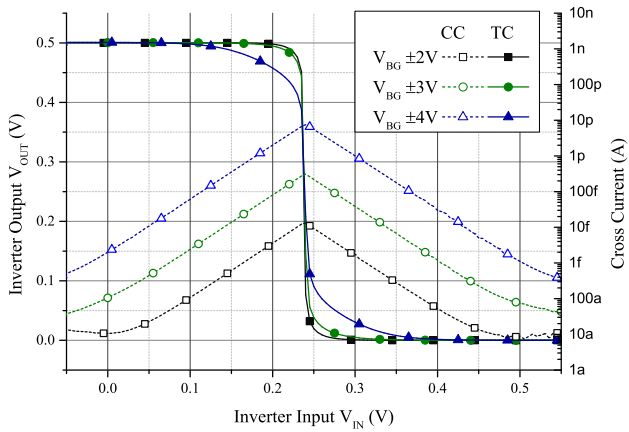

Fig. 13. Inverter transfer characteristic (TC) and cross current (CC) for dual metal front-gate device. Work function of FG metal M1 = 4.2 eV and M2 = 5.2 eV. Asymmetric supply $V_{SS}$ = 0.5 V, $V_{DD}$ = 0 V.